\documentclass[11pt]{article} 

\usepackage{mdwlist}
\usepackage{contigrefs}
\usepackage[utf8]{inputenc} 
\usepackage{subcaption}
\usepackage{caption}
\usepackage{sidecap}
\usepackage{setspace}
\usepackage{wrapfig}

\usepackage{afterpage}

\usepackage{geometry} 
\usepackage{graphicx} 

\usepackage{booktabs} 
\usepackage{array} 
\usepackage{paralist} 
\usepackage{verbatim} 
\usepackage{subfig} 
\usepackage{amsmath} 
\usepackage{amssymb} 
\usepackage{epstopdf} 
\usepackage{color}

\usepackage{fancyhdr} 
\usepackage{sectsty}
\allsectionsfont{\sffamily\mdseries\upshape} 

\DeclareMathOperator\erf{erf}

\usepackage[nottoc,notlof,notlot]{tocbibind} 
\usepackage[titles,subfigure]{tocloft} 

\title{Assortative Mating: Encounter-Network Topology and the Evolution of Attractiveness}
\author{S. Dipple$^{1,2}\footnote{E-mail: dippls@rpi.edu}$, T. Jia$^{3}$, T. Caraco$^{4}$, G. Korniss$^{1,2}$, B. K. Szymanski$^{2,5,6}$}
\date{}

\begin{document}
\maketitle

\begin{flushleft}
$^{\bf{1}}$ Department of Physics, Applied Physics, and Astronomy, Rensselaer Polytechnic Institute,
110 8$^{th}$ Street, Troy, NY, 12180-3590 USA \\
$^{\bf{2}}$ Network Science and Technology Center,
Rensselaer Polytechnic Institute, 110 8$^{th}$ Street, Troy, NY, 12180-3590 USA \\
$^{\bf{3}}$ College of Computer and Information Science, Southwest University, Chongqing, 400715, P. R. China
\\
$^{\bf{4}}$ Department of Biological Sciences, University at Albany, Albany NY 12222, USA
\\
$^{\bf{5}}$ Department of Computer Science,
Rensselaer Polytechnic Institute, 110 8$^{th}$ Street, Troy, NY, 12180-3590 USA 
\\
$^{\bf{6}}$ Wroclaw University of Science and Technology, 50-370 Wroclaw, Poland

\end{flushleft}

\section*{Abstract}
We model a social-encounter network where linked nodes match for reproduction in a manner depending probabilistically on each node's attractiveness.  
The developed model reveals that increasing either the network's mean degree or the ``choosiness'' exercised during pair-formation increases the strength of positive assortative mating.  
That is, we note that attractiveness is correlated among mated nodes.  
Their total number also increases with mean degree and selectivity during pair-formation.  
By iterating over model mapping of parents onto offspring across generations, we study the evolution of attractiveness.  
Selection mediated by exclusion from reproduction increases mean attractiveness, but is rapidly balanced by skew in the offspring distribution of highly attractive mated pairs.

\section*{Introduction}
Most animals assort positively for mating \cite{Jiang_2013}; that is, values of a phenotypic or genotypic trait correlate positively across a population's mated pairs \cite{Alpern_1999,Shine_2001}.  
The strength of assortment varies among taxonomic groups and categorical traits.  
But phenotypic similarity between paired females and males, often with respect to body size or visual signals, occurs far more often that does negative assortment or random mating \cite{Jiang_2013,Janetos_1980}.  
Humans are not an exception \cite{Berscheid_1971}.  
Mate choice in humans produces partner similarity with respect to several traits, including age, social attitudes, height and attractiveness \cite{Zietsch_2011}.  Our study focuses on attractiveness, which we invoke as a surrogate for any genetically variant, continuous trait correlating positively across pairs.

Assortative mating is sometimes adaptive.  Under disruptive selection, individuals may adaptively avoid producing lower-fitness intermediates by assorting positively for reproduction \cite{Coyne_2004}.  However, in many cases assortative mating arises as a consequence of some other ecological process. For example, if a phenotypic trait covaries spatially or temporally with habitat in both sexes, the population's spatio-temporal structure can induce assortative mating in the absence of selection on mate choice \cite{Jiang_2013}.  Importantly, trait similarity within mating pairs may drive the evolution of that trait, independently of the reason for assortative mating \cite{Lynch_1998,Bolnick_2012}.

Several disciplines, including population genetics and social psychology, have explored relationships involving individual pair-bonding preferences, assortative mating, and the evolutionary consequences of homogamy \cite{Kalick_1986,Kondrashov_1998,McPherson_2001,Bearhop_2005}.  Certain models include the realism of stochasticity in encounters between potential mates, and in both pair formation and dissociation \cite{Jia_2015,ZHJWW_2014}.  However, most available models for assortative mating assume fully-connected social structure.  That is, every female may encounter every male in the same population.  Realistically however, any individual has social contact with a limited number of potential mates, which can be characterized by the degree distribution of a bipartite network (females and males) \cite{Corviello_2012,Jia_2014}.  The network topology has a profound impact on many properties of social networks.\cite{BA_1999,WS_1998,JK_2013,SSSK_2013,KSSK_2015} 
In this case, it can govern many aspects of a system such as the number of mated nodes in a population and the strength of assortative mating across those mated nodes \cite{Jia_2015,ZHJWW_2014}.

Our analysis extends this line of inquiry.  
We assume assortative mating with respect to attractiveness, and vary the degree of ``choosiness'' or selectivity exercised during pair-formation.  
We show how the strength of assortative mating increases with an encounter network's average degree.  
We also find, contrary to intuition, that the number of mated nodes in a large population increases as this population becomes more selective during pairing, provided that the system has sufficient time to complete interactions.  
Finally, we assume that attractiveness is a heritable trait, and show how the population-level distribution of attractiveness evolves under assortative mating.

\section*{Methods}
Throughout this work, we assume that each discrete generation has the same size, and that the sex ratio is unity.  Selection acts on individuals through access to reproduction.  All successfully paired individuals have the same mean number of offspring, independently of their attractiveness; any remaining individuals leave no offspring.  This assumption lets us focus on how network topology and attractiveness interactively affect breeding inclusion \emph{vs} exclusion.

\subsection*{The Encounter Network}
We construct a bipartite graph with $2N$ nodes divided equally between subsets $A$ and $B$.  
Each node is assigned links according to a degree distribution $P(k)$; a node's links connect it to nodes of the other subset.  
The network's average degree is then $\langle k \rangle=\Sigma kP(k)$.  
Each node $A_i \in A$ has a weight $a_i \in [0,1]$ which is a continuous random variable that represents the node's attractiveness.
In this work, we initialize these variables as uniform random variables on $[0,1]$  
Nodes in subset $B$ are assigned their weights in exactly the same way.  
We then link nodes stochastically based on the degree distribution and denote the set of all links as $L$.
The links enable nodes for interaction and eventually pairing.

\subsection*{Pair-formation dynamics}
To begin, consider the pairing dynamics described in \cite{Jia_2014} (which we modify below).  
If nodes $A_i$ and $B_j$ are linked, we denote that link as $l_{i,j}=\{A_i,B_j\} \in L$ and has an associated weight defined by its endpoints as

\begin{equation}
w_{i,j}=a_ib_j.
\label{linkstr}
\end{equation}

In the model, all links are initially in the \emph{potential} state.
There are two other states of a link, temporary and permanent.
Only three transitions are possible, from potential to the temporary state and from temporary to either the potential or permanent state; the permanent state being an absorbing state.
The general flow of the pair-formation dynamics goes as follows:
\begin{enumerate} \itemsep 0pt
\item[1. ] A random link $l_{i,j}=\{A_i,B_j\}$ is chosen from set $L$.
\item[2. ] A uniform random number $r \in U(0,1)$ is generated and if $r<(w_{i,j})^{\beta}$ (where the exponent $\beta \geq 0$ controls the strength of selectivity (see next section for details)), the pairing condition is met and one of two transitions occur. 
\item[3a.] If $l_{i,j}$ is in the potential state, it transitions to a temporary state and every other link in a temporary state with one of its endpoints being $A_i$ or $B_j$ returns to the potential state.  
\item[3b.] If $l_{i,j}$ is in the temporary state, it transitions to the permanent state. 
Its endpoints, to which we will refer to as a mated pair, are then placed in the set $M$, and all of their links are removed from the graph.
\item[4. ] The simulation time is increased by $\dfrac{-\ln(q)}{|L|}$ where $q \in U(0,1)$.
\item[5. ] This repeats until subsets $A$ and $B$ are empty or contain only isolated nodes.
\end{enumerate}
Iteration of the process results in $M$ containing all mated pairs with all other nodes discarded.   

Given the pair-formation dynamics, an individual must pair before too many of its links are removed from the network, since each such removal decreases the chances to mate.  
Note that the pair-formation dynamics implies that the average attractiveness in subsets $A$ and $B$ prior to pair-formation can differ from the average attractiveness among individuals that become mated pairs.  Below we refer to this difference as the selection differential.  First, we specify how pair-formation might depend on ``choosiness,'' or partner selectivity.

\subsection*{Matching Selectivity}
The criterion for meeting the pairing condition (step 2 of the pair-formation dynamics) has been generalized compared to \cite{Jia_2014} in which $\beta =1$.  
As a convenience, we refer to $\beta$ as selectivity; as $\beta$ increases, the randomly selected link is less likely to meet the pairing condition when sampled.  
Of course, if $\beta = 0$, every sampled link meets the pairing condition, and $\beta = 1$ corresponds to the original step 2 above.  
One can mathematically show that the introduction of $\beta$ is similar to transforming the initial population to a new distribution. 
Supplementary Information presents a derivation of the continuous random variable $Y$, with realizations $w_{i,j}^{\beta} = \left( a_i ~b_j\right)^{\beta}$.

Increased selectivity should extend the time elapsing before all nodes in subsets $A$ and $B$ are removed or isolated.  
Ecologically, the time available for pair-bonding may be constrained, in which case greater selectivity might exclude more individuals from breeding \cite{Janetos_1980}.  
Our pair-formation dynamics does not take into account this constraint, but selectivity still can affect the likelihood a node of given degree forming a pair.
It is worth noting that nodes in $A$ and $B$ are assigned attractiveness and connectivity by the same random process and are drawn from the same distributions.
It follows that sets $A$ and $B$ have statistically equivilent properties of attractiveness and interconnectivity.

\subsection*{Computational Procedures}
To model a network of individuals and their encounter links, we use an Erd\H{o}s-R\'{e}nyi graph \cite{Erdos_1959}.  
The degree distribution will approximate a truncated Poisson probability function with mean $\langle k \rangle$.  
Hence nodes of high degree occur rarely.  
The graph is constructed by selecting two random nodes, one from subset $A$ and the other from subset $B$.
These nodes become linked so long as the selected nodes do not already have a link connecting them.
The created link is then added to $L$.  
This process is continued until $|L| = N\langle k \rangle$.

\subsubsection*{Rejection-free simulation}
During pair-formation, changes occur, that is the system's history is updated, only when the pairing condition is met.  Advancing the simulation so that we skip the events in which the pairing condition is not met offers a computational advantage, as when $\beta$ increases the frequency of failure to meet the pairing condition increases.  Therefore, we model pairing as the sequence of events generated by $|L|$ independent Poisson processes employing a rejection-free scheme \cite{BKL_1975,Gillespie_1976,Gilmer_1976}.

First, we construct the probability distribution for choosing which link will next meet the pairing condition.  
The probability that any given link will be chosen in step 1 of the pair formation dynamics is $\dfrac{1}{|L|}$. 
It follows that the probability that a given link $l_{i,j}$ will meet the pairing condition stated above is $\dfrac{(w_{i,j})^{\beta}}{|L|}$.  
By excluding the outcomes where the pairing condition is not met, the probability distribution for which link will meet the pairing condition is as follows.
\begin{equation}
P_{l_{i,j}}=\dfrac{ w_{i,j}^{\beta}}{\sum\limits_{l_{i',j'} \in L} w_{i',j'}^{\beta}}
\label{gillspie}
\end{equation}
\noindent
This distribution remains static until the network structure changes. 
The transition from a potential to temporary state does not impact the transition rates of any link in the system.
Only a transition to a permanent state causes the distribution to be updated.
For a rejection-free scheme, we modify the following in the pair-formation dynamics. In step 1, a link is selected by generating a uniformly distributed random number $r\in U(0,1)$ and mapping it into the inverse of the cumulative distribution function associated with Eq. \ref{gillspie} to identify the selected link $l_{i,j}$, which by definition meets the pairing condition of step 2.
Step 4 is also modified so that the time between events of meeting the pairing conditions is the random variable $\Delta T$ defined by the following equation.
\begin{equation}
\Delta T=\dfrac{-\ln(q)}{\sum\limits_{l_{i',j'} \in L} w_{i',j'}^{\beta}}
\label{gill_time}
\end{equation}
where $q$ is a uniformly distributed random number on $(0,1)$.

Our implementation serves primarily to accelerate simulation.  It also helps to explain the limiting case of $\beta \rightarrow \infty$.
Let $w_{max}$ be the maximum link weight in the population and $n_{max}\geq 1$ be the number of links with such weight. 
After dividing the numerator and denominator in Eq. \ref{gillspie} by $w^{\beta}_{max}$ in the limit of $\beta \rightarrow \infty$ the denominator $\sum_{l_{i',j'}\in L} (w_{i',j'}/w_{max})^{\beta}$ tends to $n_{max}$ because all terms corresponding to links with less than the maximum weight will be reduced to $0$.
Likewise, the numerator $(w_{i,j}/w_{max})^{\beta}$ will go to either $1$ if $w_{i,j}=w_{max}$ or $0$ otherwise.
Hence, we will have a constant probability of $1/n_{max}$ to select a link with the maximum weight. 
This choice becomes deterministic if there is only one such link. Note that when $w_{max}<1$, using this limit causes the time between each pairing conditions to become infinite. In such a case, we cannot use Eq. \ref{gill_time} to analyze the time series unless we normalize the link weights $w_{i,j}^{\beta} \rightarrow \left( \frac{w_{i,j}}{w_{max}} \right)^{\beta}$. 
This produces a very trivial time series as the denominator summation in Eq. \ref{gill_time} will tend to $n_{max}$ and become independent of which link is chosen.

\subsection*{Reproduction and Offspring Attractiveness}

Once pair-formation has ended, we use the mated pairs in set $M$ to produce the next generation of nodes. 
$A^{(g)}$ and $B^{(g)}$ represent the subsets of nodes in generation $g$, \emph{prior to} that generation's pair-formation.  
Initially by definition, we have $A^{(0)}\equiv A$ and $B^{(0)}\equiv B$.  
We define $a^{(g)}$ and $b^{(g)}$ as attractiveness values before pair-formation in generation $g$ with their initial values being $a^{(0)}\equiv a$ and $b^{(0)}=b$.  
Pair-formation in generation $g$ constructs the set $M^{(g)}$ with its attractiveness-pairs $\{{\hat{a}}^{(g)}, {\hat{b}}^{(g)}\}$. 
Thereafter, the mated pairs in set $M^{(g)}$ produce the next generation's $A^{(g+1)}$ and $B^{(g+1)}$.  
We assume that any offspring of a given pair in $M^{(g)}$ has attractiveness $x$ sampled from a truncated normal density with moments conditioned on the parents' attractiveness levels.  
That is, offspring attractiveness has the conditional probability density:
\begin{equation}
f(x|\mu ,\sigma )=\dfrac{2 e^{-\dfrac{1}{2}\left( \dfrac{x-\mu}{\sigma} \right)^2} }{\sigma \left( \erf{ \left( \dfrac{\mu}{\sigma \sqrt{2}} \right)} + \erf{  \left( \dfrac{1-\mu}{\sigma \sqrt{2}} \right) } \right) } ; x\in [0,1]
\end{equation}

\noindent
Reproduction proceeds for $N$ steps as follows.
\begin{enumerate} \itemsep 0pt
\item Randomly choose a mated pair $\{A_i^{(g)}, B_j^{(g)}\}$ in set $M^{(g)}$.
\item Generate a node in each of the subsets $A^{(g+1)}$ and $B^{(g+1)}$; for each node generated, sample its attractiveness from $f(x|\mu ,\sigma )$ with $\mu =\dfrac{{\hat{a}_i}^{(g)} + {\hat{b}_j}^{(g)}}{2}$ and $\sigma =G\vert \mu - {\hat{a}_i}^{(g)}\vert$ where $G$ is the offspring variance with $G > 0$.
\end{enumerate}
\noindent
After all nodes have been generated, we assign links in a similar manner as their parents (using the same distribution and parameters), but completely independently.
This process also preserves the statistically similar properities that a given generation has in terms of attractiveness and interconnectivity between sets $A^{(g)}$ and $B^{(g)}$.

\begin{figure}[t]
\centerline{\includegraphics[width=\textwidth]{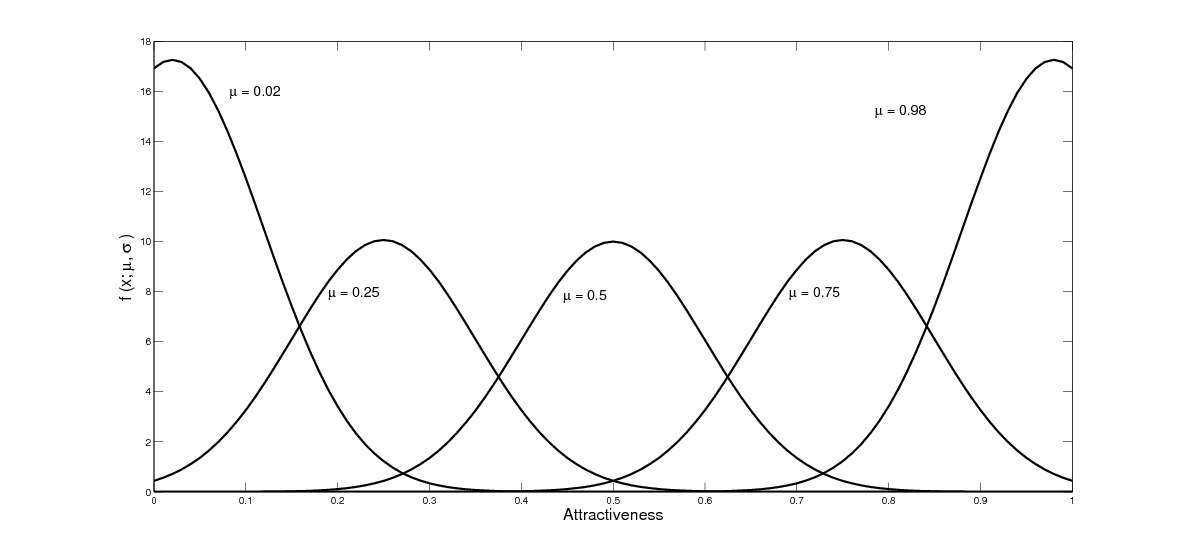}}
\caption{\textbf{Offspring Phenotype Distribution.}  For mid-parents $\mu = 0.25,~0.5,~0.75$, the distribution of offspring attractiveness is symmetric about the mid-parent, as usually assumed in quantitative  genetics.  However, for $\mu = 0.98$ ($\mu = 0.02$) most offspring have attractiveness phenotypes less than (more than) the mid-parent.  $\sigma = 0.1$.
}
\label{xpdf}
\end{figure}

Several assumptions introduced here require elaboration.  
We restrict attractiveness to the unit interval, whereas quantitative genetic models commonly treat them as unbounded phenotypic traits \cite{Slatkin_1976}.   
This can be overcome by transforming the attractiveness distribution to one which is boundless.  
This transformation must be in line with the probabilities of meeting the pairing condition in the intended system.  
Bounds on attractiveness induce an obvious complication under assortative mating.  
Given a mated pair $\{{\hat{a}}^{(g)}, {\hat{b}}^{(g)}\}$, both attractiveness values can be close to unity.  
The upper bound on $x$ implies that their offspring will likely be less attractive than the midparent $\mu$, since $f(x|\mu ,\sigma )$ has negative skew. (Figure \ref{xpdf})  Similarly, the converse is also true when the mated pair has $\mu$ close to zero. The offspring distribution near the parental-phenotype boundaries does not strongly affect our analysis of attractiveness evolution, since we focus on population change at interior values.

We assume that the mean and the (approximate) variance of the offspring trait distribution depend on the parents' phenotypes.  
Most quantitative genetic models assume a reproductive variance independent of the parents' trait values \cite{Slatkin_1976}.  
Ultimately, we anticipate that phenotypic variability among a given pair's offspring will increase with the difference between parental trait values.

\subsection*{Selection: Differential and Response}
In our model, the selection differential $S_g$ is the difference between mean attractiveness among individuals of generation $g$ that pair for reproduction and mean attractiveness among members of the same generation at birth.  
$S_g \neq 0$ implies that average attractiveness differs between individuals who attract a mate and those that do not.  
We have:
\begin{equation}
S_g = \frac{\sum_i {\hat{a}_i}^{(g)} + \sum_j {\hat{b}_j}^{(g)} }{2 \vert M^{(g)}\vert } - \frac{\sum_i {a_i}^{(g)} + \sum_j {b_j}^{(g)}}{2N}
\end{equation}
\noindent
Our model's response to selection $R_g$ is the difference between mean attractiveness among individuals of generation $(g + 1)$ at birth and mean attractiveness among their parents at birth.  
We have:
\begin{equation}
R_g = \left[ \sum_i {a_i}^{(g+1)} + \sum_j {b_j}^{(g+1)} - \sum_i {a_i}^{(g)} - \sum_j {b_j}^{(g)} \right]/2N
\end{equation}
\noindent
By definition, the population evolves when $R_g \neq 0$.

\section*{Results}

\subsection*{Assortative Mating: Selectivity and Degree}
First, we consider the distribution of mated pairs $\{\hat{a}, \hat{b}\}$ formed during simulation, as selectivity $\beta$ and average number of links per node $\langle k \rangle$ are varied.  
Each simulation included $2N = 2 \times 10^4$ nodes.  
Using the same initial conditions (uniformly distributed attractiveness values), we conducted 20 simulations and averaged results.

\begin{figure}[t]
	\centering
		\includegraphics[width=\textwidth]{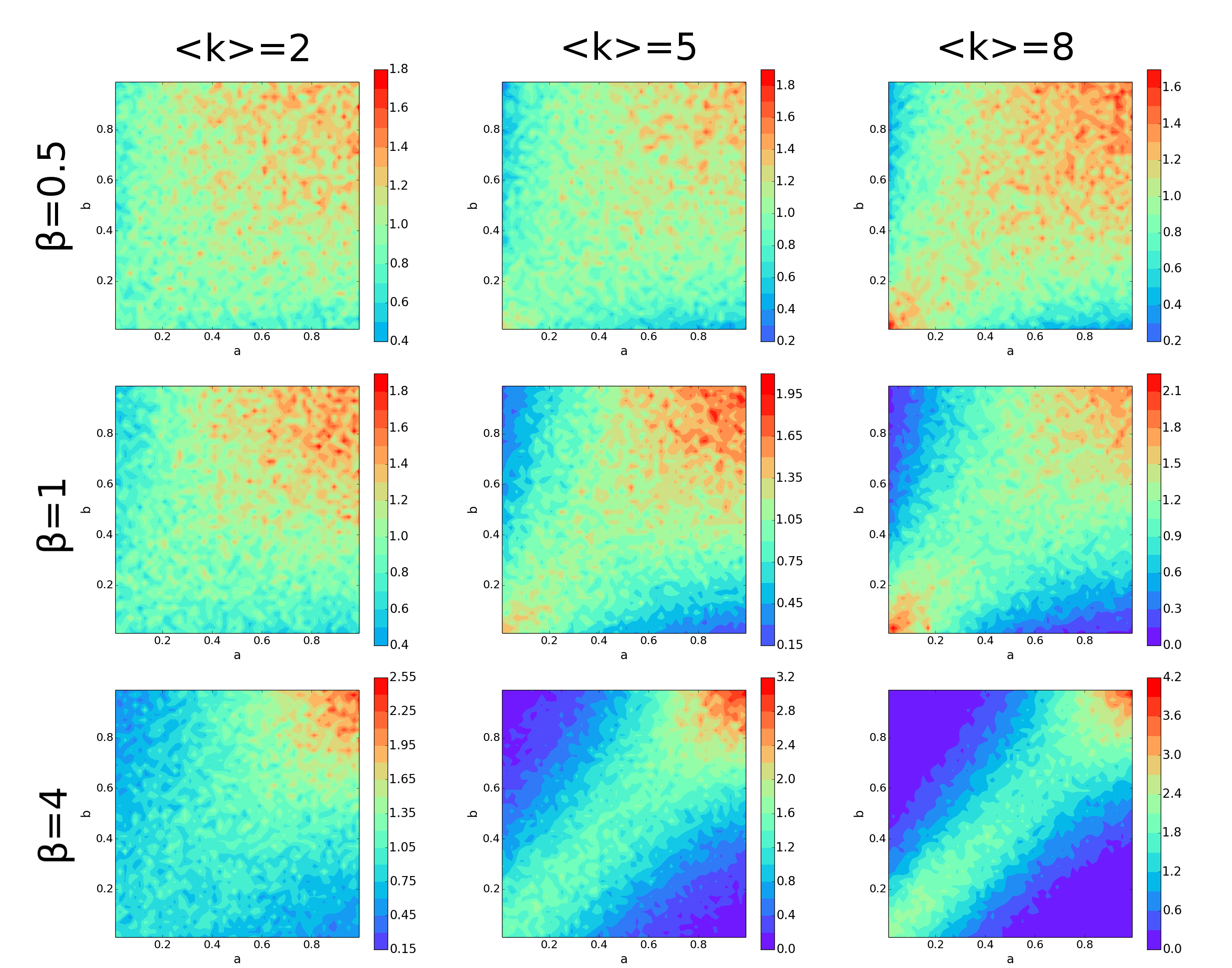}
	\caption{\textbf{Attractiveness Joint Probability Distribution.} Each mated pair is binned into a 0.02 by 0.02 bin according to the pair's attractiveness set.
	Each distribution is normalized and then averaged, and the bins are smoothed for visual purposes.
	The top row corresponds to $\beta = 0.5$, middle row $\beta = 1$, and bottom row $\beta =4$.
	The left column corresponds to $\langle k \rangle = 2$, middle column $\langle k \rangle = 5$, and right column $\langle k \rangle = 8$.}
	\label{gen_1_dists}
\end{figure}

Figure \ref{gen_1_dists} shows the relative frequencies of attractiveness values for mated pairs.  
Fixing $\langle k \rangle$, increasing selectivity promotes the strength of assortative mating.  
Fixing $\beta$ while increasing the network's average degree increases the strength of assortment.  
For $\beta < 1$, the effect of greater average degree is relatively small, since encounters so readily lead to pairing.  
Indeed, the combination of low $\beta$ and small $\langle k \rangle$ (upper left panel) generates mated pairs with attractiveness close to uniformly distributed over all levels.  
High selectivity and large degree (lower right panel) nearly eliminate bonding of mated pairs with a large difference in attractiveness, and matching is highly assortative.  
These results on pair-formation accord with intuition, and motivate us to apply the model to the evolution of attractiveness.

Note that in each panel of Figure \ref{gen_1_dists}, the most frequent mated pairs involve two highly attractive individuals. 
As described above, these matches tend to form earlier during the pair-formation process; highly attractive individuals are unlikely to be excluded from reproduction.

\subsection*{Number of Mated Pairs Formed}
\begin{figure}[!t]
	\centering
		\includegraphics[width=\textwidth]{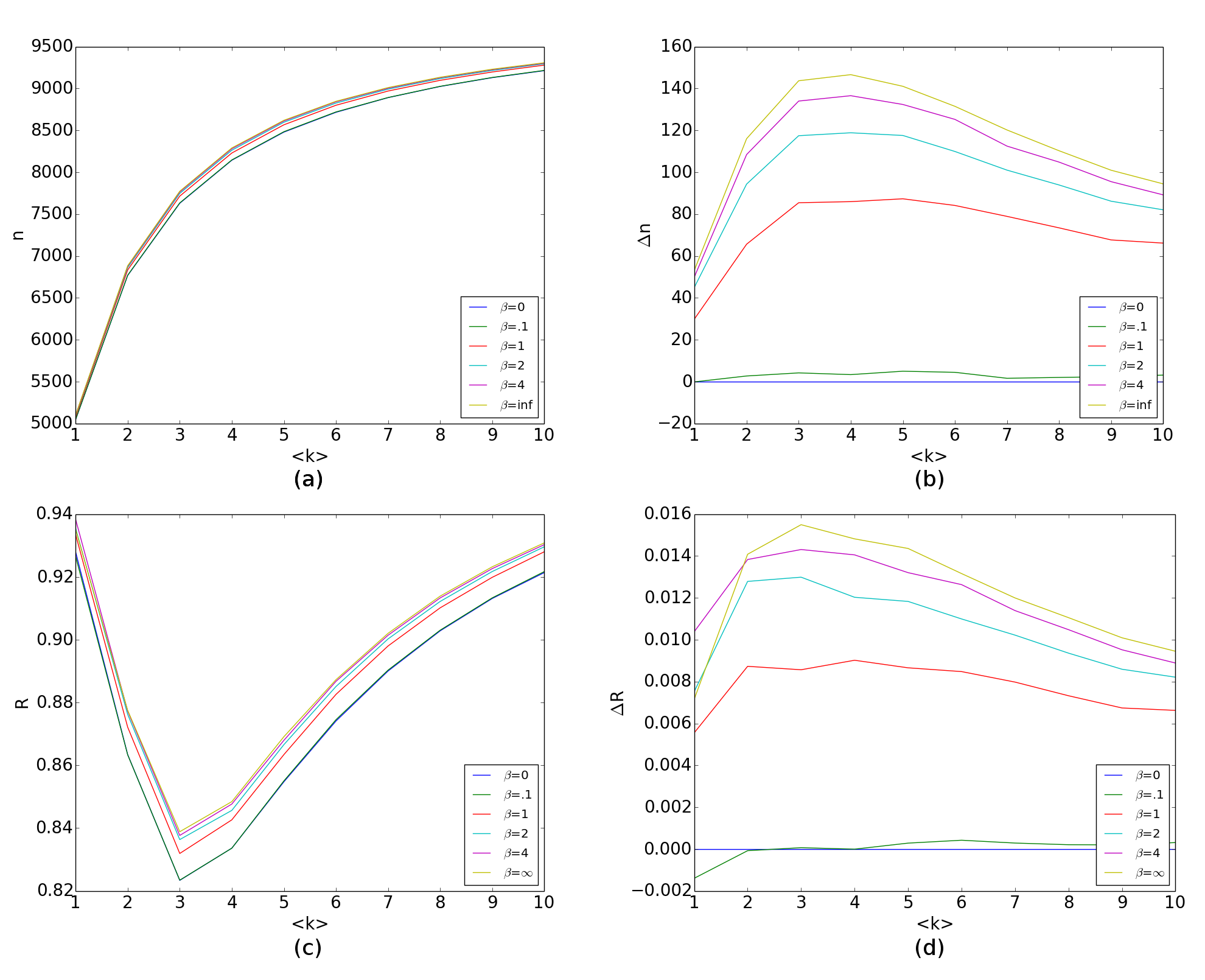}
	\caption{\textbf{Number of Matches and Matching Efficiency.} 1000 realizations are produced for various values of average node degree and selectivity.
	(a) The number of mated pairs $n$.
	(b) The relative changes in the number of matches is defined as $\Delta n=n(<k>,\beta )-n(<k>,0)$.
	(c) The matching efficiency $R$, which is the average ratio of the number of matches for a given system over the maximum possible number of matches for that same system.
	(d) The relative changes in the matching efficiency is defined as $\Delta R=R(<k>,\beta )-R(<k>,0)$.
	A one-to-one correspondence is not present between (b) and (d) due to each realization of the matching having its own ratio and that ratio being averaged, rather than dividing the average number of matches by the average maximum number of matches.}
	\label{max_matching}
\end{figure}

The total number of mated pairs formed, $\vert M \vert = n$, increases in a decelerating manner as the mean node degree increases (Figure \ref{max_matching}a).  
Not surprisingly, increasing the total number of feasible encounters results in fewer individuals excluded from reproduction.

Fixing $\langle k \rangle$, we find that greater selectivity results in an increased number of mated pairs (Figure \ref{max_matching}b).  
The effect of selectivity is strongest in networks of low average degree $(\langle k \rangle = 3, 4)$.  
Interestingly, the ratio of $n$ to the maximal number of mated pairs that \emph{could} form (in the same network) is minimal where the effect of selectivity on $n$ is maximal (Figure \ref{max_matching}c, \ref{max_matching}d).  
That is, where network topology results in the greatest proportional exclusion of individuals from mating, the increase in pairing due to greater selectivity attains a maximum.  
Averaged over attractiveness, increased selectivity during pair-formation reduces the likelihood that an individual will be excluded from mating.

To explain this observation, Figure S1 shows that nodes of low degree, averaged over attractiveness, have a greater probability of becoming apart of a mated pair as $\beta \rightarrow \infty$.  
This is because increasing selectivity increases the strength of assortative mating, and that mated pairs of mutually high-attractiveness form earliest for any $(\langle k \rangle,~\beta)$-combination.  
Figure S1 suggests that greater selectivity causes nodes with low degree, but high attractiveness to have an increased chance of becoming a mated pair, decreasing the number of links removed when this happens.
Fewer nodes of intermediate (or low) attractiveness then need be excluded from mating due to more links being available, therefore $n$ increases.



\subsection*{Time to Match}

\begin{figure}[t]
	\centering
		\includegraphics[width=.5\textwidth]{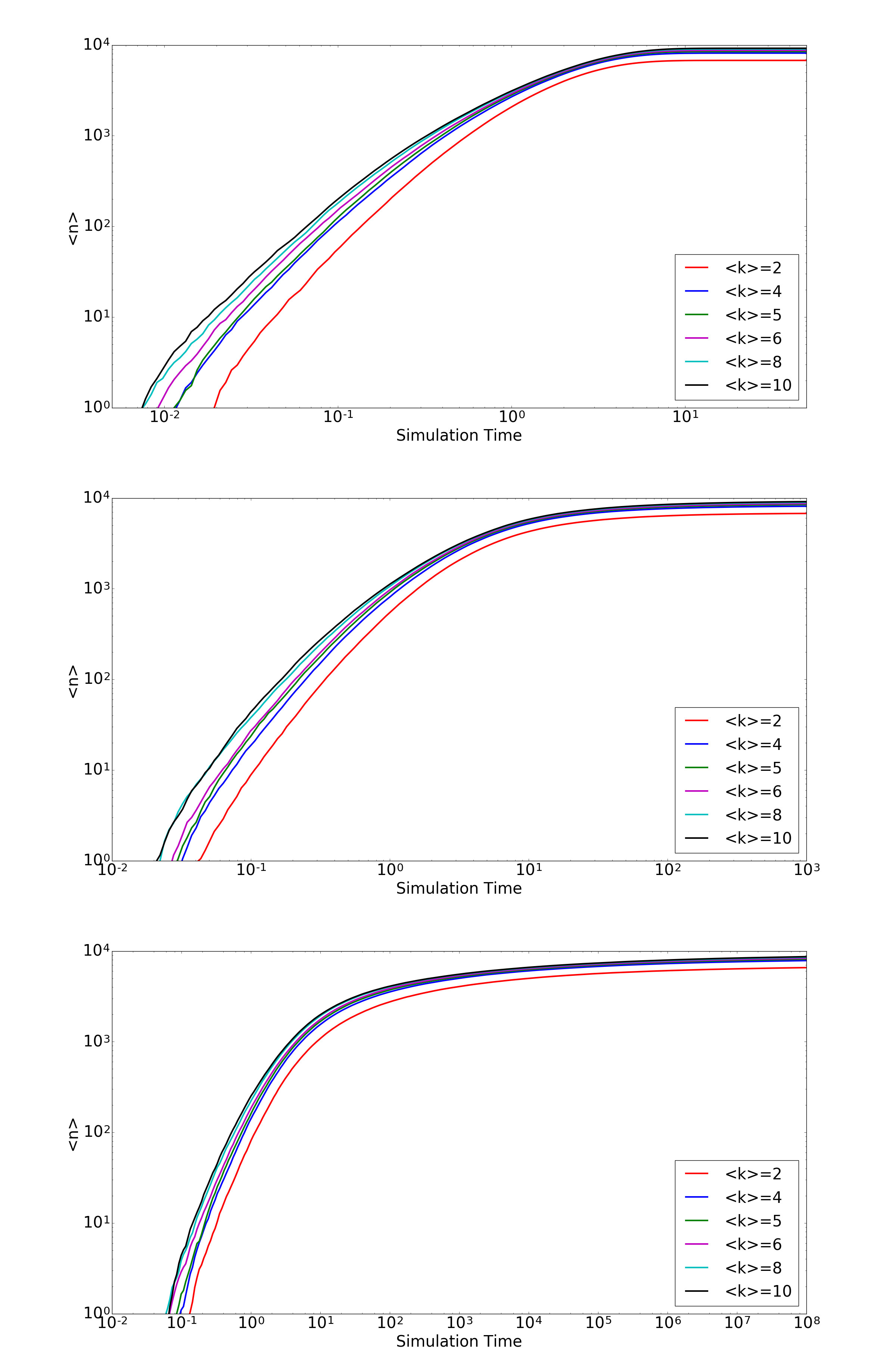}
	
	\caption{\textbf{Time to Match I.} The average number of matches $\langle n \rangle$ vs time averaged over twenty trials.
	The top figure corresponds to a constant $\beta =0.1$, middle $\beta =1$, and bottom $\beta =4$.}
	\label{time_b}
\end{figure}

\begin{figure}[t]
	\centering
		\includegraphics[width=\textwidth]{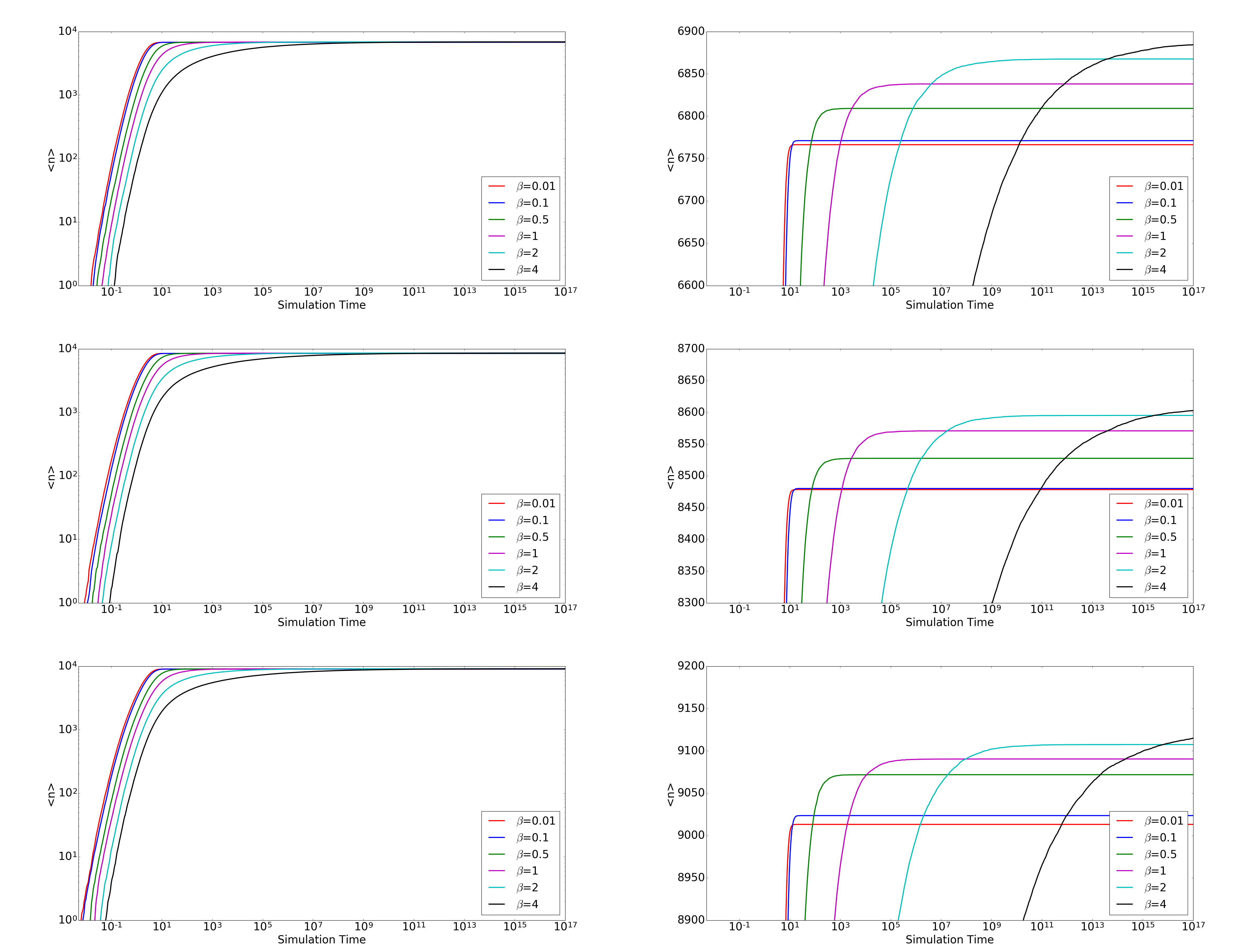}
	
	\caption{\textbf{Time to Match II.} The average number of matches $\langle n \rangle$ vs time averaged over twenty trials.
	The top row corresponds to a constant $\langle k \rangle =2$, middle row $\langle k \rangle =5$, and bottom row $\langle k \rangle =8$.
	The right column magnifies the left column regions that contain intersections of functions.
	}
	\label{time_k}
\end{figure}

In the model, simulations do not have an upper bound on how long individuals are allowed to match.
In general, real life situations have an upper bound on how long interactions are allowed to take place.
Figure \ref{time_b} shows the number of matches as a function of time for various $<k>$ at constant $\beta$.
Figure \ref{time_k} shows similar plots, but for various $\beta$ at constant $<k>$.

We observe that for a given average node degree, the time required to match increases by orders of magnitude as selectivity increases.
This is expected as increased selectivity in general increases the time between successful pairing conditions.
As shown above with sufficient time higher selectivities eventually catch up to lower selectivity in terms of the number of matches.
In addition there is an overhead time before any system can begin producing mated pairs which exists in all realizations.
This is due to the courtship mechanism where all links need to meet the pairing condition at least twice.
Because there is a low chance to randomly select a given link, it on average takes a significant amount of time for the first mated pair to form.

\section*{Attractiveness Evolution}
To address phenotypic evolution, we first fix selectivity $\beta$, and vary the network's mean degree. 
Figure S2 shows how the bivariate distribution of mated pairs changes from an initial to post third round of pair-formation, with $\beta = 1$.  
The distribution of mated pair attractiveness levels becomes more condensed after each generation in each case.  
The dispersion of the distribution looks similar for various $\langle k \rangle$, but is statistically different.  
Since assortative mating increases with $\langle k \rangle$, the distribution's small dependence on $\langle k \rangle$ suggests that the distribution of phenotypes at birth rapidly changes from generation to generation.

Figure S2 might suggest bias towards phenotypic values near the center of the distribution.  
But the result merely reflects the below average chance that individuals of high attractiveness will be excluded from reproduction (directional selection), and the negative skew of the offspring distribution of high-attractiveness parents - so that their offspring are of intermediate attractiveness.

\begin{figure}[tp]
	\centering
		\includegraphics[width=.8\textwidth]{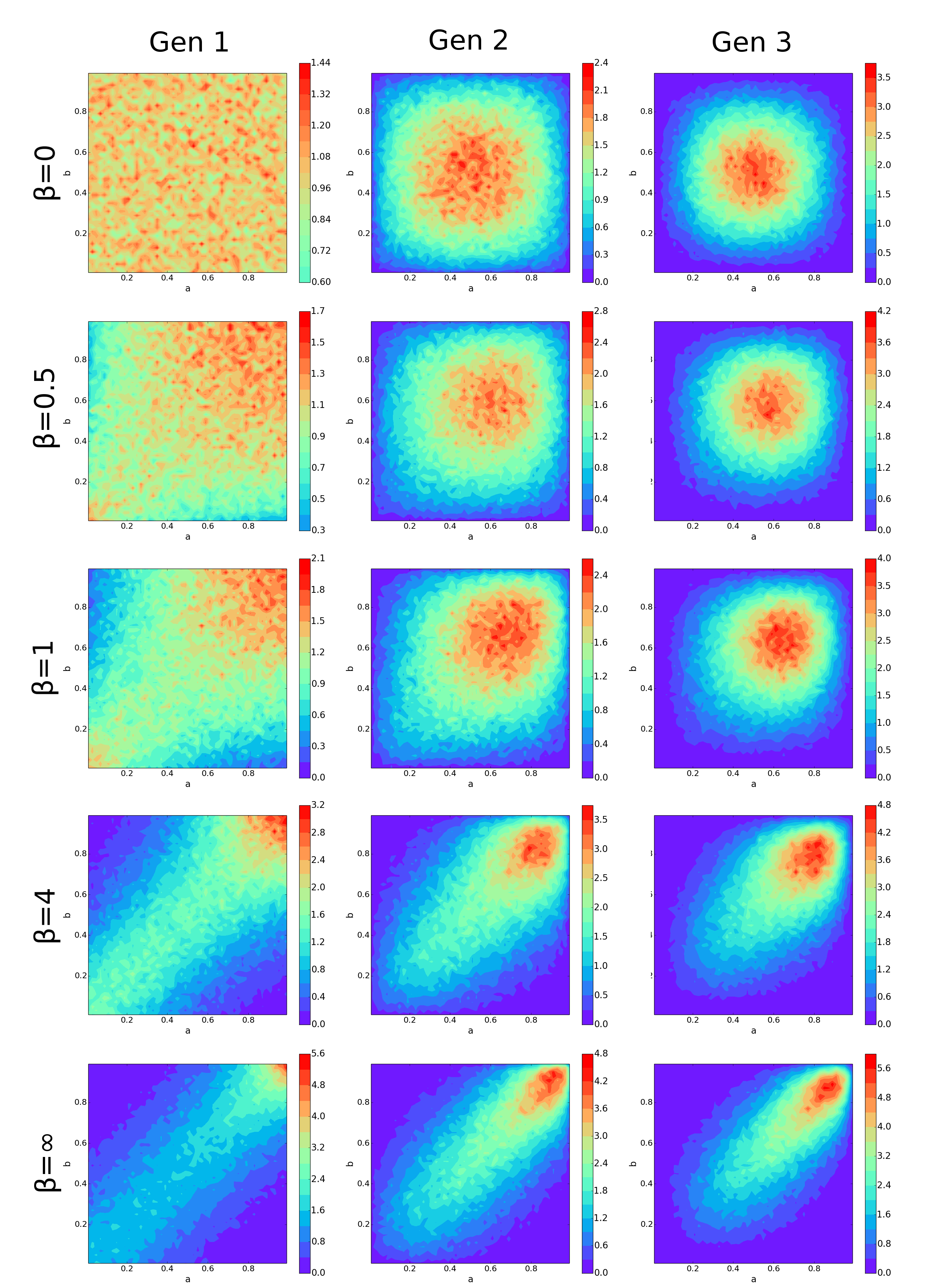}
	
	\caption{\textbf{Attractiveness Joint-Probability Distribution Evolution.} Each mated pair is binned into a 0.02 by 0.02 bin according to the pair's attractiveness set.
	Each distribution is normalized and then averaged, and the bins are then smoothed for visual purposes.
	From the top row going down, the rows use $\beta =0$, $\beta =0.5$, $\beta =1$, $\beta =4$, and $\beta =\infty$.
	The left column corresponds to generation zero, middle column generation one, and right column generation two.
	All distributions were generated with $\langle k\rangle =5$ and $G=0.75$.}
	\label{gen_dists_beta}
\end{figure}

Fixing $\langle k \rangle$, we show effects of increasing selectivity in Figure \ref{gen_dists_beta}.  
Again, the bivariate distribution of mated pairs begins to converge quickly.  
For $\beta < 1$, the distribution's radial symmetry (after only three generations) suggests a statistical loss of assortative mating.  
However, for $\beta > 1$, the distribution regains positive assortment, and the most common mated pairs are between two highly attractive nodes - a consequence of the underlying evolution of the trait to a higher mean attractiveness. 
The mean of the distribution also increases with increasing $\beta$.

We briefly note that increasing $G$, which increases the variance of the offspring-trait distribution for any mated pair $\{\hat(a), \hat(b)\}$, increases the dispersion of the bivariate distribution of mated pairs (Figure S3).

Iterating the selective pair-formation process and subsequent reproduction will drive the population to statistical equilibrium; that is, $R_g \rightarrow 0$ as $g \rightarrow \infty$.  
At or near equilibrium, the expected increase in mean attractiveness per generation balances the decrease due to the negative skew in the offspring-production distribution $f(x|\mu ,\sigma )$.

\begin{figure}[!t]
	\centering
		\includegraphics[width=\textwidth]{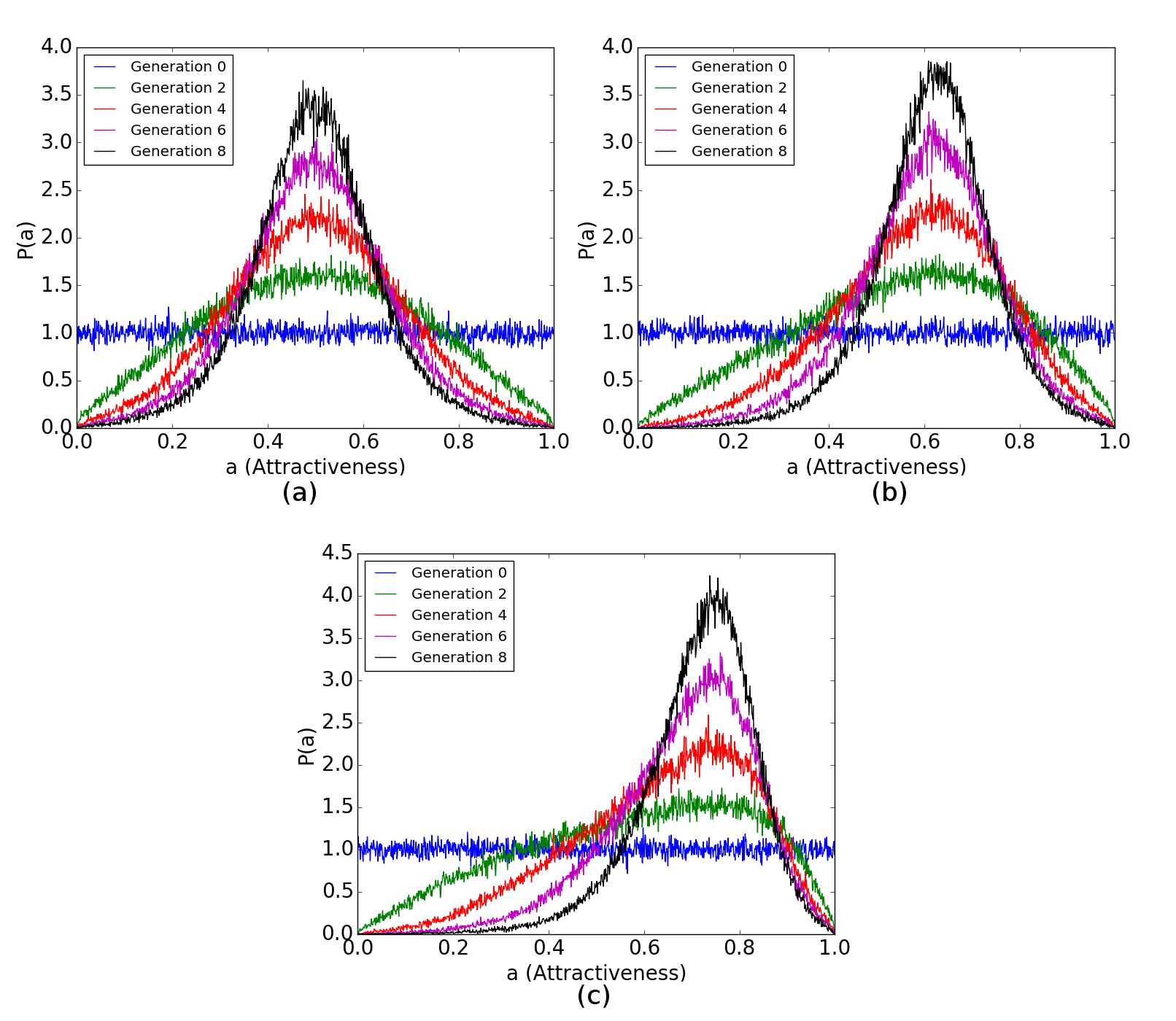}

	\caption{\textbf{Attractiveness Distribution Evolution.} Sampled probability density $P(a)$ for a node to have attractiveness $a$. 
	$P(a)$ is sampled only from the set $A$ of nodes (because $A$ and $B$ are statistically equivalent) for each generation then binned with a bin size of 0.001.
	The distribution is then averaged over twenty realizations.
	(a) $\beta =0$.
	(b) $\beta =1$.
	(c) $\beta =4$.		
	All distributions correspond to $\langle k \rangle = 5$ and $G=1$.}
	\label{fit_dists}
\end{figure}

Figure \ref{fit_dists} shows a temporal series of univariate attractiveness distributions for several different selectivity values.  
Mean attractiveness increases with $\beta$; recall Figure \ref{gen_dists_beta}. 
The variance of the distribution appears independent of selectivity, since dispersion depends strongly on $G$.

To find the long term behavior, we look at 100 generations for various parameter values. 
The mean and variance of these distributions can be seen in Figure S5. 
Both metrics reach a statistical equilibrium after a sufficient number of generations.

\subsection*{Offspring Number of Matches}
The number of matches is examined as a function of generation.
Figure S4 shows an interesting decline in the number of matches.
Because all other factors known to change the number of matches are constant, this indicates that the number of matches is dependent on the attractiveness distribution of the population.

This can be caused by the input distributions for the matching process approaching a similar value.
As mentioned above, selectivity can also mathematically be thought as modifying the input distribution for the matching process.
For $\beta =0$, the input distribution is a singularity at one, which corresponds to all individuals possessing the same attractiveness.
For later generations, the distributions are narrowing.
This approach to distributions where all nodes possess the same attractiveness is likely causing this decrease as there are too few low degree, high attractiveness nodes to increase the number of matches.

\section*{Discussion}
The first set of results presented here addresses variation in assortative mating with respect to attractiveness.  
Increasing either the encounter network's mean degree or pair-formation selectivity increases assortment across mated pairs.  
Associated with these intuitive results, we find that the earliest and most frequent mated pairs involve two highly attractive individuals.

Our results indicate that increasing mean degree increases the number of mated pairs formed within a generation.  
Counter-intuitively, we find that increased selectivity increase the number of mated pairs.  
Hence, fewer individuals are  excluded from reproduction when the pair-formation dynamics exhibit greater selectivity.  
The latter effect is associated, statistically, with nodes of low degree.

The second set of results investigate evolution of mean attractiveness when mated pairs assort positively.  
Selection mediated by exclusion from breeding can increase mean attractiveness.  
Reproductive variance led to production of enough moderate phenotypes to balance any selective advantage of high attractiveness.  
For $\beta < 1$ (weak selectivity) evolution over several generations reduced assortative mating.  
However, for $\beta > 1$ (strong selectivity) the evolving population retained positive assortment by attractiveness.


\section*{Acknowledgments}
This research was supported in part
by the Office of Naval Research Grant No.  N00014-15-1-2640, and
the Army Research Laboratory under Cooperative Agreement Number W911NF-09-2-0053 (the ARL Network Science CTA).
T. Jia is also supported by the Natural Science Foundation of China (No. 6160309).
B. K. Szymanski is also supported by the European Commission under the 7th Framework Programme, Agreement Number 316097, and by the Polish National Science Centre, the decision no. DEC-2013/09/B/ST6/02317.
The views and conclusions contained in this document are those of the authors and should not be interpreted as
representing the official policies, either expressed or implied, of the Army Research Laboratory or the US Government.

\section*{Author Contributions}
B.K.S. conceived the research;
S.D., T.J., T.C., G.K., and B.K.S. designed the research;
S.D. and T.J. implemented and performed numerical experiments and simulations;
S.D., T.J., T.C., G.K., and B.K.S. analyzed data and discussed results;
S.D., T.J., T.C., G.K., and B.K.S. wrote, reviewed, and revised the manuscript.

\section*{Additional Information}
Competing financial interests: The authors declare no competing financial interests.

\newpage


\begin{thebibliography}{100}




\bibitem{Jiang_2013}
Jiang Y, Bolnick DI, Kirkpatrick M (2013)
Assortative mating in animals.
American Naturalist 181:E125--E138.

\bibitem{Alpern_1999}
Alpern S, Reyniers D (1999)
Strategic mating with homotypic preferences.
Journal of Theoretical Biology 198:71--88.

\bibitem{Shine_2001}
Shine R, O'Connor D, LeMaster M, Mason R (2001)
Pick on someone your own size: ontogenetic shifts in mate choice by male garter snakes results in size-assortative mating.
Animal Behavior 61:1133--1141.

\bibitem{Janetos_1980}
Janetos AC (1980)
Strategies of female choice: a theoretical analysis.
Behavioral Ecology \& Sociobiology 7:107--112.

\bibitem{Berscheid_1971}
Berscheid E, Dion K, Walster E, Walster GW (1971)
Physical attractiveness and dating choice: a test of the matching hypothesis.
Journal of Experimental Social Psychology 7:173--189.

\bibitem{Zietsch_2011}
Zietsch BP, Verweij KJH, Heath AC, Martin NG (2011)
Variation in human mate choice: simultaneously investigating heritability, parental influence, sexual imprinting, and assortative mating.
American Naturalist 177:605--616.

\bibitem{Coyne_2004}
Coyne J, Orr H (2004)
Speciation. Sinauer, Sunderland, MA.

\bibitem{Lynch_1998}
Lynch M, Walsh B (1998)
Genetics and Analysis of Quantitative Traits. Sinauer, Sunderland, MA.

\bibitem{Bolnick_2012}
Bolnick DI, Kirkpatrick M (2012)
The relationship between intraspecific assortative mating and reproductive isolation between divergent populations.
Current Zoology 58:484--492.

\bibitem{Kalick_1986}
Kalick SM, Hamilton TE (1986)
The matching hypothesis reexamined.
Journal of Social Psychology 51:673--682.

\bibitem{Kondrashov_1998}
Kondrashov AS, Spaak M (1988)
On the origin of species by means of assortative mating.
Proceedings of the Royal Society B: Biological Sciences 265:2273--2278.

\bibitem{McPherson_2001}
McPherson M, Smith-Lovin L, Cook JM (2001)
Birds of a feather: homophily in social networks.
Annual Review of Sociology 27:415--444.

\bibitem{Bearhop_2005}
Bearhop S, Fieldler W, Furness RW, Votier SC, Waldron S, Newton J, Bowen GJ, Berthold P, Farnsworth K (2005)
Assortative mating as a mechanism for rapid evolution of a migratory divide.
Science 310:502--504.

\bibitem{ZHJWW_2014}
Zhou B, He Z, Jiang LL, Wang NX, Wang BH (2014) 
Bidirectional selection between two classes in complex social networks. Sci Rep 4.

\bibitem{Jia_2015}
Jia T, Spivey RF, Szymanski B, Korniss G (2015)
An analysis of the matching hypothesis in networks.
PLoS One 10:e0129804.

\bibitem{Corviello_2012}
Corviello L, Franceschetti M, McCubbins MD, Paturi R, Vattani A (2012)
Human matching behavior in social networks: an algorithmic perspective.
PLoS One 7:e41900.

\bibitem{Jia_2014}
Jia T, Spivey R, Korniss G, Szymanski B (2014)
A network approach in analysis of the matching hypothesis.
Bull. American Physical Society 59:B16.


\bibitem{BKL_1975}
Bortz AB, Kalos MH, and Lebowitz JL,
A New Algorithm for Monte Carlo Simulations of Ising Spin Systems.
J. Comput. Phys. {\bf 17}, 10--18 (1975).

\bibitem{Gillespie_1976}
Gillespie DT,
A general method for numerically simulating the stochastic time evolution of coupled chemical reactions.
J. Comput. Phys. {\bf 22}, 403--434 (1976).

\bibitem{Gilmer_1976}
Gilmer GH,
Growth on Imperfect Crystal Faces.
J. Crystal Growth {\bf 35}, 15--28 (1976).


\bibitem{Erdos_1959}
Erd\H{o}s,  P, \& R\'enyi, A (1959)
On random graphs.
Publ. Math. Debrecen 6:290--297.

\bibitem{Slatkin_1976}
Slatkin M, Lande R (1976)
Niche width in a fluctuating environment.
American Naturalist 110:31--55.

\bibitem{prod_rand}
Springer, M D (1979).
The Algebra of Random Variables.
Wiley.

\bibitem{BA_1999}
Albert-László Barabási, Réka Albert
Emergence of Scaling in Random Networks
Science: 509-512 (1999)

\bibitem{WS_1998}
Duncan J. Watts, Steven H. Strogatz
Collective dynamics of 'small-world' networks
Nature 393, 440-442, doi:10.1038/30918 (1998)

\bibitem{JK_2013}
Jia T, Kulkarni RV. 
On the structural properties of small-world networks with range-limited shortcut links. 
Physica A. 2013;392(23):6118–6124. doi: 10.1016/j.physa.2013.07.060. 

\bibitem{SSSK_2013}
P. Singh, S. Sreenivasan, B. K. Szymanski, G. Korniss
Threshold-limited spreading in social networks with multiple initiators
Scientific Reports 3, Article number: 2330 (2013)

\bibitem{KSSK_2015}
P. D. Karampourniotis, S. Sreenivasan, B. K. Szymanski, G. Korniss
The Impact of Heterogeneous Thresholds on Social Contagion with Multiple Initiators
PLoS ONE 10(11): e0143020. doi:10.1371/journal.pone.0143020 (2015)


%
%
%
%
%
%
%
%
%
%
%


\end{thebibliography}
\end{document}